\documentclass[aps,prl,twocolumn]{revtex4}
\usepackage{bm,color,amsmath,amssymb,mathrsfs,latexsym,graphicx,psfrag}

\newcommand{\bk}{{\mathbf k}}

\newcommand{\be}{\begin{equation}}
\newcommand{\ee}{\end{equation}}
\newcommand{\mM}{\mathcal{M}}

\def\bea{\begin{eqnarray}}
\def\eea{\end{eqnarray}}

\begin{document}

\title{New classes of topological crystalline insulators with unpinned surface Dirac cones}
\author{Chen Fang}
\affiliation{Massachusetts Institute of Technology, Cambridge MA 02139, USA}
\author{Liang Fu}
\affiliation{Massachusetts Institute of Technology, Cambridge MA 02139, USA}

\begin{abstract}{{We theoretically predict two new classes of three-dimensional topological crystalline insulators (TCIs), which have an odd number of unpinned surface Dirac cones protected by crystal symmetries. The first class is protected by a single glide plane symmetry; the second class is protected by a composition of a twofold rotation and time-reversal symmetry. Both classes of TCIs are characterized by a quantized $\pi$ Berry phase associated with surface states and a $Z_2$ topological invariant associated with the bulk bands. In the presence of disorder, these TCI surface states are protected against localization by the average crystal symmetries, and exhibit critical conductivity in the universality class of the quantum Hall plateau transition. These new TCIs exist in time-reversal-breaking systems with or without spin-orbital coupling, and their material realizations are discussed.}}
\end{abstract}
\maketitle

The notion of symmetry protected topological (SPT) phases has recently emerged from studies on topological insulators and is now being intensively studied\cite{Schnyder2008,Hasan2010,Qi2011,Chen2012,Senthil2014}. SPT phases generically have gapless boundary states that are stable against perturbations, provided that certain symmetry is preserved. The topological property of these boundary states depends crucially on the underlying symmetry. In the well-known example of three-dimensional (3D) topological insulators, time reversal symmetry protects an odd number of surface Dirac points that are pinned to time-reversal invariant momenta (TRIM).

Recent theoretical studies\cite{Teo2008,Mong2010,Fu2011,Hsieh2012,Fang2012,Fang2013,Chiu2013,Morimoto2013,Liu2014,Alexandradinata2014,Shiozaki2014,Lu2014a,Zhang2014} have found a variety of SPT phases that are protected by crystal symmetries, termed topological crystalline insulators (TCIs)\cite{Fu2011}. A universal property of TCI phases is the presence of protected surface states on symmetry-preserving surfaces. However, depending on the underlying crystal symmetry, surface state properties of different classes of TCIs can vary significantly. In spin-rotationally-invariant TCIs protected by rotational symmetries of the crystal (such as $C_4$)\cite{Fu2011,Alexandradinata2014}, surface states exhibit band crossings that are pinned to certain high-symmetry point(s) in the SBZ. In spin-orbit-coupled TCIs protected by mirror symmetry, as realized in the SnTe class of IV-VI semiconductors\cite{Hsieh2012,Xu2012,Tanaka2012,Dziawa2012,Ando2015}, surface states exhibit Dirac points on a specific mirror-symmetric line in the SBZ, corresponding to the projection of the 2D plane with a nonzero mirror Chern number in the Brillouin zone\cite{Teo2008}.

In this work, we theoretically predict two new classes of 3D TCIs, protected by a glide plane symmetry and a space-time inversion symmetry respectively. Unlike all topological insulating phases known so far, their surface states consist of a {\it single} (more generally an odd number), {\it unpinned} Dirac point with a quantized $\pi$-Berry phase. Importantly, these new TCI phases are robust against either magnetic or nonmagnetic impurities, which by definition preserve the crystal symmetry on average. Remarkably, the disordered surface realizes, without any tuning, a critical phase in the universality class of quantum Hall plateau transition.

{\it TCI with glide symmetry} The first class of TCI exists in 3D systems (with or without spin-orbital coupling) that have a glide plane symmetry, i.e., a combination of reflection and a translation by half a lattice vector:
\bea\label{eq:glidedef}
M_G:\;(x,y,z)\rightarrow(x,y,-z)+\mathbf{a}_1/2,
\eea
where $(x,y,z)$ is the position vector and $\mathbf{a}_{1,2,3}$ are the basis of lattice vectors, out of which $\mathbf{a}_1$ is inside the $xy$-plane. A key difference between a mirror plane and a glide plane is that the mirror plane squares to identity (up to a Berry phase associated with a $2\pi$ rotation), while the glide mirror squares to a lattice translation:
\bea\label{eq:glidesquare}
M^2_G=(-1)^f{T}_{\mathbf{a}_1}.
\eea
Here $f=0$ applies to spin-rotationally-invariant systems, where reflection does not involve spin; $f=1$ applies to spin-orbit-coupled systems, where reflection acts on spin $s=\frac{1}{2}$ and squares to $-1$.

\begin{figure*}[tbp]
\includegraphics[width=0.9\textwidth]{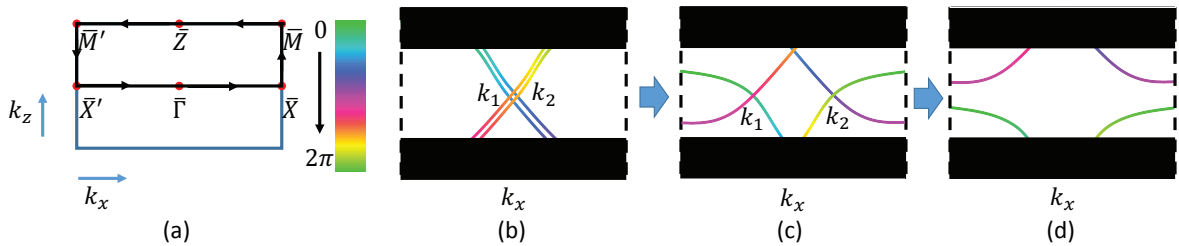}
\caption{(a) The SBZ of the $xz$-surface of a 3D (spinless) system with glide plane defined in Eq.(\ref{eq:glidedef}). (b,c,d) are schematics of the process where two identical Dirac points on a single mirror symmetric line annihilate each other through finite surface perturbation. Here in (b) we set $k_{1,2}$ slightly away from each other only to indicate that there are two instead of one Dirac points. The color bar on the left of (b) shows the color code for the phase of the glide plane eigenvalues.}
\label{fig:BZandFlow}
\end{figure*}

We now show how a glide plane can protect a crossing point in the surface bands. First, a symmetry-preserving surface must  be (i) perpendicular to the glide plane ($xy$-plane in this case) and (ii) invariant under  the translation along $\mathbf{a}_1$. Without loss of generality, we choose $\mathbf{a}_{1,2}$ to be along $x,y$-axes, respectively. The only surface that satisfies both conditions is then the $xz$-plane, whose SBZ  is plotted in Fig.~\ref{fig:BZandFlow}(a). Due to the translational symmetry in the $xz$-plane, the Hamiltonian with an open surface is diagonal in Bloch basis with crystal momentum $(k_x, k_z)$. The corresponding Bloch Hamiltonian is denoted by $h(k_x,k_z)$.
The presence of glide plane symmetry implies:
\bea\label{eq:glidesym}
\mathcal{M}_G(k_x)h(k_x,k_z)\mathcal{M}^{-1}_G(k_x)=h(k_x,-k_z).
\eea
Here the operator $\mM_G(k_x)$ represents the action of $M_G$ in Bloch basis. Note that unlike point group symmetry operators, $\mM_G(k_x)$ is a function of $k_x$. This results from Eq.(\ref{eq:glidesquare}), which implies eigenvalues of the glide plane are not constants but depend continuously on electron's momentum.

It follows from Eq.(\ref{eq:glidesym}) that all bands on the two high-symmetry lines $k_z=0$ and $\pi$ (where lattice constants are taken to be unity) can be labeled by the eigenvalues of $\mM_G(k_x)$. Using Eq.(\ref{eq:glidesquare}), and taking into account that at $T_{\mathbf{a}_1}=e^{ik_x}$, the two eigenvalues of $\mM_G(k_x)$ are
\bea\label{eq:glideeig}
m_\pm(k_x)=\pm{}i^fe^{ik_x/2}.
\eea
These eigenvalues divide the bands along $k_z=0,\pi$ into two branches that have glide plane eigenvalue of $m_+(k_x)$ and $m_-(k_x)$ respectively---hereafter referred to as $m_+$ and $m_-$ bands. Since any hybridization between a $m_+$ band and a $m_-$ band breaks the glide plane, a single crossing point $\bk_0$ between them at any momentum on the line $k_z=0$ or $k_z=\pi$ is protected. The $k\cdot p$ Hamiltonian at $\bk_0$ takes the form (up to a unitary basis change): $H (\bk)= v_x k_x \sigma_1 + v_y k_y \sigma_2$, where $\bk$ denotes the momentum relative to the Dirac point; and the action of glide mirror on the two degenerate surface states at $\bk_0$  is represented by $M_G (\bk_0) \propto \sigma_1$, up to a $U(1)$ phase factor. Provided that glide plane symmetry is preserved,  perturbations can shift the band crossing point $\bk_0$ along the high-symmetry line, but cannot open a gap. This leads to a symmetry-protected surface Dirac point that is not pinned to a specific point. Furthermore, the stability of this Dirac point has a topological origin arising from the quantization of the Berry's phase, a point to which we will return later.

Having addressed the protection of a single surface Dirac cone, we proceed to study the stability of two Dirac cones, each of which is locally protected by $M_G$. In this case, can we adiabatically tune the Hamiltonian to fully gap the surface without closing the bulk gap? This turns out to be a subtle issue that has not been encountered in topological insulating phases studied so far. To answer this question, two cases should be discussed separately: (i) the two Dirac points appear on the same line, either $k_z=0$ or $k_z=\pi$ and (ii) there is one Dirac point on each mirror invariant line.

As an example of the first case, we combine two identical copies of the aforementioned TCI with a single surface Dirac cone. The combined system then has two surface band crossings that appear at the same point on same mirror symmetric line in SBZ, say $(k_0, 0)$ [Fig.~\ref{fig:BZandFlow}(b)]. Infinitesimal perturbation on the surface cannot fully gap the spectrum, because the two right-going (left-going) modes have the \emph{same} mirror eigenvalue. Two Dirac points are hence `locally stable'. However, we find the corresponding surface state spectrum can be and can only be gapped out by sufficiently strong deformations, i.e., it is globally unstable. To see this, we need to use a key property of the glide plane symmetry, as indicated in Eq.(\ref{eq:glideeig}): a $m_+$-band is connected with a $m_-$-band at the BZ boundary at $k_x=\pm\pi$, because the phase factor on the right hand side gives an additional minus sign when $k_x$ goes to $k_x+2\pi$. Consider a finite surface perturbation that pushes $k_1$ to the left and $k_2$ right [Fig.\ref{fig:BZandFlow}(c)]. When they meet each other again at the SBZ boundary, according to the above property, the right-going (left-going) modes have opposite mirror eigenvalues so $k_{1,2}$ can annihilate each other [Fig.\ref{fig:BZandFlow}(d)]. We emphasize that both the local stability and global instability of two Dirac points are key characteristics associated with the glide plane symmetry, which have not appeared elsewhere. We further note that since two Dirac points can only annihilate each other by crossing the SBZ boundary, the new TCI phase with glide mirror symmetry
cannot be treated in continuum models where $k$-space is effectively a sphere rather than a torus. Generally, the torus nature of SBZ must be considered in studying nonsymmorphic symmetries of a lattice.

\begin{figure}[tbp]
\includegraphics[width=0.45\textwidth]{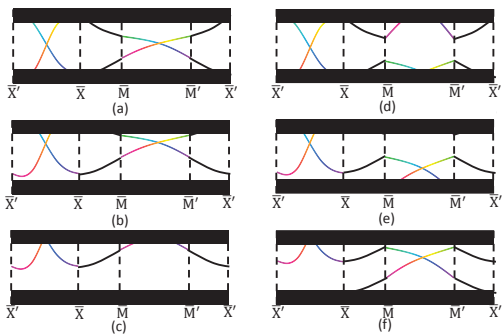}
\caption{(a-c) show the process of gapping the spectral flow when there are two band crossings, on $\bar{X}'\bar\Gamma\bar{X}$ and $\bar{M}'\bar\Gamma\bar{M}$ respectively. (d-f) show the continuous flow when there is only one band crossing, and how the crossing moves from $\bar{X}'\bar\Gamma\bar{X}$ to $\bar{M}'\bar\Gamma\bar{M}$.}
\label{fig:Flow}
\end{figure}

In the second case, we consider the spectral flow of the band dispersion in the SBZ along the path $\bar{X'}\bar\Gamma\bar{X}\bar{M}\bar{Z}\bar{M'}\bar{X'}$, shown in Fig.~\ref{fig:BZandFlow}(a) by the arrows. Fig.~\ref{fig:Flow}(a) and Fig.\ref{fig:Flow}(c) show typical spectral flows for a trivial and a nontrivial phase, respectively. We need the following principle for the analysis: along $\bar{X}'\bar\Gamma\bar{X}$ and $\bar{M}'\bar{Z}\bar{M}$, the bands must appear in pairs that cross each other. The proof of the principle is given in Sec. I of Supplementary Materials (also see Ref.[\onlinecite{Parameswaran2013}]). In Fig.~\ref{fig:Flow}(a), there are two band crossings on $\bar{X}'\bar\Gamma\bar{X}$ and $\bar{M}'\bar\Gamma\bar{M}$, respectively. When the chemical potential on the surface increases, the two bands move upward in energy together, as in Fig.~\ref{fig:Flow}(b) and are eventually pushed into the conduction bands as in Fig.~\ref{fig:Flow}(c), leaving a full gap on the surface. In Fig.~\ref{fig:Flow}(d), there is one band crossing along $\bar{X}'\bar\Gamma\bar{X}$, and no crossing along $\bar{M}'\bar{Z}\bar{M}$. When the surface chemical potential increases, the two bands move upward in energy together. However, since bands must appear in crossing pairs along $k_z=0$ and $k_z=\pi$, pushing up the chemical potential will `pull out' a pair of bands from the valence bands, as shown in Fig.~\ref{fig:Flow}(e). As as result, the Dirac point on $\bar{X}'\bar\Gamma\bar{X}$ moves to $\bar{M}'\bar{Z}\bar{M}$ without closing the bulk gap, and the flow remains [see Fig.~\ref{fig:Flow}(f)].

We have now shown that a single surface Dirac cone can be protected by a glide mirror symmetry, but not two cones. Due to the bulk-edge correspondence, this implies the existence of a $Z_2$ topological invariant for the bulk. The analytic and explicit expression of the invariant involve the theoretical tool of non-Abelian Berry phase and Wilson loop\cite{Wilson1974,Yu2011,Alexandradinata2014a,Taherinejad2014,Alexandradinata2014b}, which we leave to Sec. II of Supplementary Materials. There, we also provide a lattice model as an example of this TCI phase in a spinless system.

To summarize, the TCI with a glide mirror symmetry has a single Dirac point that (i) is unpinned to any high-symmetry point, and (ii)  under symmetry perturbations that does not close the bulk gap, can move along two mirror invariant lines as well as shift between the two lines. Can we have a new TCI phase with a single Dirac point that can freely move in the SBZ without being confined to any high-symmetry point or line in the SBZ? Below we provide an affirmative answer to this question.

{\it TCI with space-time inversion symmetry} We now show that an anti-unitary symmetry $C_2*T$, the combination of a twofold rotation and time-reversal, protects a $Z_2$ TCI phase with a single Dirac cone on the surface perpendicular to the twofold axis, whose location in the SBZ is completely unpinned. This symmetry is hereafter referred to as a space-time inversion symmetry, as $C_2*T$ operation on the surface sends $(x,y,t)$ to $(-x,-y,-t)$.

We first note the identity $(C_2*T)^2= I$, which holds for both systems with and without spin-orbit coupling, because both $C_2^2$ and $T^2$ are equal to $-I$ for the former, and $I$ for the latter. This identity allows us to represent the action of $C_2 *T$ on the two degenerate states at a (putative) Dirac point by $C_2*T=K\sigma_1$,  where $K$ is complex conjugation.  Then, one can easily verify that the Dirac Hamiltonian $H(\bk)= v_x k_x\sigma_1+ v_y k_y\sigma_2$  is indeed compatible with $C_2*T$ symmetry. Perturbations that are invariant under $C_2 * T$ correspond to $\sigma_1$ or $\sigma_2$, which simply shift the Dirac point. The Dirac mass term $\sigma_3$ is odd under $C_2 *T$ and hence forbidden. This simple analysis demonstrates the protection of an unpinned Dirac point by $C_2 * T$.

We now further show that two cones can annihilate each other under perturbations preserving the $C_2*T$ symmetry. The Hamiltonian for two identical copies of surface Dirac cones has the following form $H=\tau_0\otimes(k_x\sigma_x+k_y\sigma_y)$, with $C_2*T=K\tau_0\otimes\sigma_1$, where $\tau_{0,1,2,3}$ are identity and Pauli matrices acting on the species space. A symmetry preserving perturbation $\delta{m}\tau_2\otimes\sigma_3$ can gap the whole spectrum. This implies the $Z_2$ nature of the TCIs protected by $C_2*T$ symmetry, with an odd number of surface Dirac points. The $Z_2$ topological classification can also be derived from analysis of the bulk topology, as we show in Sec. III of Supplementary Materials.

The  protected surface states of TCIs with space-time inversion symmetry are characterized by a quantized $\pi$ Berry phase, for any loop enclosing all Dirac points thereof. As stated above, the same is true for TCIs with the glide mirror symmetry, provided that the loop enclosing the Dirac point is symmetric. Here we provide a general proof for both cases. First we show that the Berry's phase is independent of the geometry (such as size or shape) of the loop. According to Stokes theorem, the difference between Berry phases associated with two loops mod $2\pi$ is given by the integral of Berry curvature over the region bounded by them. For systems with glide plane symmetry, the Berry's curvature for the surface states satisfies $F(k_x,k_z)=-F(k_x,-k_z)$, due to the oddness of $F$ under reflection $z \rightarrow -z$.  So for any loop symmetric about a mirror invariant line, the Berry's curvature integral inside the loop subtract the singular band crossing point vanishes. For systems with space-time inversion symmetry, Berry's curvature satisfies $F(k_x,k_y)=-F(k_x,k_y)$ due to the oddness of $F$ under time-reversal symmetry, and hence vanishes everywhere in $\bk$ space. So the Berry's curvature integral is again zero. Therefore, only Dirac points make a singular contribution to the Berry phase. As is well-known, the Berry phase of each Dirac point is $\pi$. Therefore, an odd number of Dirac points on the TCI surface guarantees a quantized $\pi$ Berry's phase, for both glide mirror or space-time inversion symmetry. (The quantization of $\pi$ Berry phase by $C_2*T$ and its implication for TCI phase were mentioned in a recent work\cite{Serbyn2014};  The local stability of a Dirac point in the presence of $C_2*T$ symmetry was also noted\cite{Hsieh2013,Fang2013a}.)

\textit{Material realizations} The two new classes of $Z_2$ TCI can be realized in systems with or without spin-orbit coupling. They are discussed separately below.

In spinful systems, the two new classes of $Z_2$ TCI are consistent with time-reversal symmetry (TRS). One can simply consider a $Z_2$ strong topological insulator which also has glide mirror and/or twofold rotation symmetry. Pick a surface that preserves the glide plane or the twofold axis. It is guaranteed to possess a single Dirac cone at one TRIM,  protected by spinful TRS. Now let us add perturbations that break TRS but preserve the glide plane or $C_2*T$;  we immediately obtain the $Z_2$ TCI phases found in this work. This observation makes it very simple to find these new TCIs in spinful systems. For example, the newly discovered topological Kondo insulator SmB$_6$\cite{Dzero2012,Lu2013,Jiang2013,Li2014} has $C_2$-symmetry on the $(001)$-surface. Adding any magnetic field parallel to the surface preserves $C_2*T$. Hence SmB$_6$ with an in-plane field can be considered as a $Z_2$ TCI protected by $C_2*T$, having an odd number of Dirac cones located away from any TRIM.

To realize  the new TCI phases in systems without spin-orbit coupling (or equivalently, spinless systems) requires breaking TRS. This can be shown by proving that  a single Dirac cone characteristic of these TCIs cannot appear on the surface of a time-reversal-invariant spinless insulator. To see this, first note that Dirac cones at non-TRIM  must appear in pairs with opposite momenta. This leaves the possibility of having a single Dirac cone at one TRIM.
However, given that time-reversal symmetry is represented by $T=K$ (up to a gauge) for spinless fermions and it reverses $\bk$ measured from the TRIM, a 2D Dirac Hamiltonian is simply not allowed by $T$. This is because only one of the three Pauli matrices, namely, $\sigma_2$, is reversed under $K$, while a 2D Dirac Hamiltonian such as $H(\bk) = v_x k_x \sigma_1 + v_y k_y \sigma_2 $ or any other form must involve two Pauli matrices. This concludes that a single Dirac cone cannot exist on the surface of a spinless system with time-reversal symmetry.

Therefore, the key requirement for spinless TCIs is to break TRS while preserving the relevant crystal symmetry of either glide mirror or space-time inversion. One may search in magnetic insulators with negligible SOC, since many types of magnetic order have at least one of the two symmetries. Alternatively, we note that photonic crystals may be a very promising platform for finding these TCI phases, because their structures and crystal symmetries can be easily manipulated\cite{Lu2013a,Lu2014}.

\textit{Stability against disorder and quantum Hall criticality} We now show that topological surface states of TCIs with either glide mirror or $C_2*T$ symmetry are fully robust against any type of disorder (magnetic or nonmagnetic) and cannot be exponentially localized even under strong disorder on the surface. Similar to the case of a disordered TI surface with random magnetic impurities\cite{Fu2012}, the delocalization of TCI surface states here is protected by the average symmetry\cite{Hsieh2012,Ringel2012,Mong2012,Liu2012,Fulga2014,Ando2015} -- by definition the ensemble of TCIs with all disorder realizations must preserve the relevant symmetry. Consider a finite area on the surface where surface states are gapped by disorder. The mass $m$ acquired by the Dirac cone can either be positive or negative, corresponding to two types of domains. The symmetry (glide mirror or $C_2*T$) maps a $m>0$ domain to a $m<0$ domain and vice versa. Importantly, the two domains have quantized half-integer Hall conductivity $\frac{e^2}{2h}$ and $-\frac{e^2}{2h}$ respectively; between the two domains of opposite signs there exists a chiral edge mode. Therefore, in the presence of random disorder, these chiral modes percolate through the surface and the surface states remain delocalized. This argument also shows that the disordered TCI surface is symmetry-enforced to be in a critical phase (rather than fine-tuned to a critical point) in the same universality class as the quantum Hall plateau transition. Therefore, the surface exhibits a universal longitudinal conductivity on the order of $\frac{e^2}{h}$, and a nontrivial scaling of longitudinal and Hall conductivity as a function of an applied out-of-plane magnetic field $B$, which gaps the Dirac point and drives each surface into a quantum Hall state with a Hall conductance ${\rm sign}(B) e^2/2h$.  The space-time inversion symmetry protects the gapless surface Dirac point by forbidding this out-of-plane magnetic field, thereby defining a TCI phase. Other symmetries such as reflection or any \emph{improper} rotation can also forbid the perpendicular field, and therefore guarantees at least a $Z_2$ classification of TCIs\cite{Fang2012}, as in the two classes studied here.

To conclude, we have theoretically predicted two new classes of 3D $Z_2$ TCI that have unpinned surface Dirac cones, which are protected by a glide plane and $C_2*T$, respectively. The $Z_2$ nature distinguishes the new TCI from the $Z$ TCI protected by mirror symmetry such as SnTe. Because of the `unpinned' nature of the surface Dirac cone, the TCI phases studied in this work lie beyond the previous classification of TCIs with nonsymmorphic space groups based solely on considerations of 2D irreducible representations of the wallpaper groups\cite{Liu2014}. Besides looking at the surface states, we also mathematically prove the $Z_2$ classification which directly reveals the bulk band topology. We emphasize that the new TCI phases can be realized in both spinful and in spinless systems. In spinful systems, these TCIs can be realized by applying perturbations that break TRS yet preserving glide plane or $C_2*T$. The realization of TCIs in spinless systems is an interesting subject that we leave to future work.

We thank Ling Lu and Timothy Hsieh for collaborations on related works. This work was supported by the STC Center for Integrated Quantum Materials, NSF Grant No. DMR-1231319 (CF), and the DOE Office of Basic Energy Sciences, Division of Materials Sciences and Engineering under Award No. DE-SC0010526 (LF).

\onecolumngrid
\section{Proof of a principle for the band crossings in the presence of a glide plane symmetry}

In the text, we mention that along a mirror invariant line in the SBZ, all edge modes must appear in pairs that cross each other, in the absence of chiral modes. We use $k\in[0,\pi)$ to parameterize this line. All edge modes along this line can be separated into $m_+$-bands and $m_-$-bands, and in the main text, we have proved that
\bea
E_{+i}(k=2\pi)=E_{-i'}(k=0),
\eea
where $E_{\pm{i}}(k)$ denotes the dispersion of the $i$-th band in the $m_\pm$ sector. If $E_{+i}(k)$ does \emph{not} cross any other bands, then due to
\bea
E_{+i}(k=0)\neq{E}_{+i}(k=2\pi),
\eea
$E_{+i}(k)$ is a chiral mode, contradicting the assumption of the absence of chiral modes. Therefore, there must be one other band that crosses $E_{+i}(k)$. But since only a $m_-$-band can cross a $m_+$-band, $E_{+i}(k)$ must cross some $E_{-j}(k)$. The principle is proved.

\section{Explicit expression for the $Z_2$ invariant for 3D insulators with a glide plan symmetry}

In the main text, we mention that we have found the explicit expression for the $Z_2$ invariant using Wilson loops.
\begin{figure}
\label{fig:BZ}
\includegraphics[width=10cm]{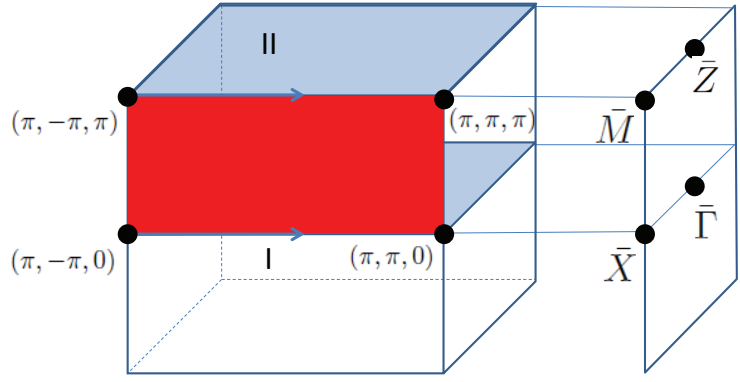}
\caption{The 3D Brillouin zone and the surface Brillouin zone on the $(010)$-surface. The blue planes are where the mirror Berry curvatures are evaluated in Eq.(\ref{eq:nI},\ref{eq:nII}) and the red (half) plane is where the total Berry curvature is evaluated in Eq.(\ref{eq:n0}). The two Wilson loops are marked with arrows.}
\end{figure}

We define two Wilson loops: (I) a straight line from $(\pi,0,0)$ to $(\pi,2\pi,0)$ and (II) a straight line from $(\pi,0,\pi)$ to $(\pi,2\pi,\pi)$. (See the two lines denoted by `I' and `II' in Fig.\ref{fig:BZ}.) On these lines one can separate the occupied spaces into two, one spanned by states with mirror eigenvalue $+i^{F-1}$ and the other with mirror eigenvalue $-i^{F-1}$. The eigenvalues of the Wilson loop, which are gauge independent, are denoted by $\lambda^{I,II}_{j,\pm}$. Then we consider the plane defined by $k_z=0$ and $k_z=\pi$. On these planes, the bands can be diagonalized into blocks having mirror eigenvalue $m_\pm(k_x)\equiv\pm{e}^{ik_x/2}$. We can define the Berry phases associated with the two blocks to be $B_{\pm}^{I,II}$ for the two planes. For now we would assume that the Chern number on both planes are zero, so $B_+^{I,II}=-B_-^{I,II}$. Next we consider a plane that has the two loops as its boundary (the red half plane in Fig.(\ref{fig:BZ}). The Berry curvature integral on this plane is denoted by $B_0$.

We can now define three integers. First we observe that on the plane $k_z=0$, any $m_+$ state at $k_x=-\pi$ continuously evolve to a $m_-$ state at $k_x=\pi$. This is equivalent to saying that all the eigenvalues corresponding to $m=+i$ of loop I must evolve into the eigenvalues corresponding to $m=-i$ by going through the $k_z=0$ plane. At the same time, we know that these eigenvalues gain a total phase of $B_+^I$ in the process. Therefore, we have
\bea\label{eq:nI}
\sum_{j=1,...,N_{occ}/2}\log(\lambda^{I}_{j+})+B^{I}_+=\sum_{j=1,...,N_{occ}/2}\log(\lambda^{I}_{j-})+2n_I\pi.
\eea
Similarly on $k_z=\pi$, we can define $n_{II}$:
\bea\label{eq:nII}
\sum_{j=1,...,N_{occ}/2}\log(\lambda^{II}_{j+})+B^{II}_+=\sum_{j=1,...,N_{occ}/2}\log(\lambda^{II}_{j-})+2n_{II}\pi.
\eea
Finally, all eigenvalues of loop I must evolve into all the eigenvalues of loop II by going through the red half plane defined above. At the same time, we know that these eigenvalues gain a total phase of $B_0$. Therefore
\bea\label{eq:n0}
\sum_{j=1,...,N_{occ}/2}[\log(\lambda^{I}_{j+})+\log(\lambda^{I}_{j-})]+B_0=\sum_{j=1,...,N_{occ}/2}[\log(\lambda^{II}_{j+})+\log(\lambda^{II}_{j-})]+2n_0\pi.
\eea
Finally, we can write down the topological invariant
\bea\label{eq:z2}
z_2=(n_{I}+n_{II}+n_0)\;\textrm{mod}\;2.
\eea
In order to call this number as a new quantum $Z_2$ number, we still need to establish that (i) it is gauge invariant, (ii) it does not change under gapped, continuous deformation of the Hamiltonian and (iii) there is at least one example that realizes the nontrivial phase, i.e., $z_2=1$.

(i) It is trivial to see the gauge invariance because it is defined in terms of Berry curvature integrals and Wilson loop eigenvalues, both of which are gauge invariant.

(ii) Upon a gapped, continuous deformation of the Hamiltonian, the Wilson loop eigenvalues change continuously. However, due to the logarithms used in Eq.(\ref{eq:nI},\ref{eq:nII},\ref{eq:n0}), we should still prove that the $z_2$ is unchanged even if any of the eigenvalues crosses the branch cut. We assume that the branch cut is made such that the logarithm takes value with $[0,2\pi)$. Suppose one of the eigenvalues $\lambda^{I}_{j+}$ increases its phase that it crosses the cut from below, then it is easy to see that $n_I\rightarrow{n}_I+1$ and $n_0\rightarrow{n}_0+1$ simultaneously, so $z_2$ is unchanged.

\begin{figure}[tbp]
\includegraphics[width=0.45\textwidth]{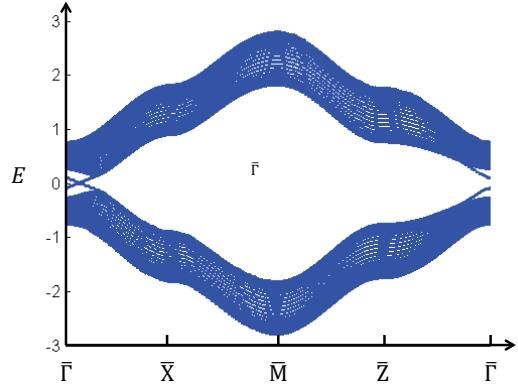}
\caption{The surface band structure for the $xz$-surface of the system given in Eq.(\ref{eq:TB1}) for the parameter set $m=2.5$, $t_0=t'_0=t''_0=t=t'={t''}=1$ and $\phi=0.4$.}
\label{fig:glidemodel}
\end{figure}

(iii) The tight-binding Hamiltonian is given by
\bea\label{eq:TB1}
H(\bk)&=&(m-t_0\cos{k_x}-t'_0\cos{k_y}-{t''}_0\cos{k_z})\Sigma_{03}
+t\sin[(k_x-\phi)/2][\sin{k_x/2}\Sigma_{11}+\cos{k_x/2}\Sigma_{21}]\\
\nonumber&+&t'\sin{k_y}\Sigma_{02}+{t''}\sin{k_z}\Sigma_{31},
\eea
where $\Sigma_{ij}=\sigma_i\otimes\sigma_j$. The glide mirror operator is given by
\bea
\mathcal{M}_G(k_x)=e^{-ik_x/2}[\cos(k_x/2)\Sigma_{10}+\sin(k_x/2)\Sigma_{20}].
\eea
It can be easily verified that
\bea
\mathcal{M}^2_G&=&e^{-ik_x},\\
\nonumber
\mathcal{M}_G(k_x+2\pi)&=&\mathcal{M}(k_x),\\
\nonumber
\mathcal{M}_GH(k_x,k_y,k_z)\mathcal{M}_G^{-1}&=&H(k_x,k_y,-k_z).
\eea
Fig.~\ref{fig:glidemodel} shows the surface states on the $(010)$-surface for a certain set of parameters (see caption). There is a single Dirac cone in the SBZ along $\bar{X'}\bar\Gamma\bar{X}$, which is consistent with the nontrivial spectral flow discussed above. This model, with a certain set of parameters, realizes a spinless $Z_2$ TCI protected by glide plane symmetry.

\section{Proof of the $Z_2$ classification for 3D insulators with $C_2*T$ from bulk topology}

In any tight-binding model, the matrix representation of $C_2*T$ is $KU$, where $U$ is some unitary matrix. Since $(C_2*T)^2=I$ for both spinless and spinful fermions, $U$ must also be symmetric, thus being orthogonal. We can hence choose another basis, which is related to the original basis by matrix $O$, such that $O^TUO=I$. In this new basis, $C_2*T$ is complex conjugation, or, $C_2*T=K$. $C_2*T$ then ensures that $H(k_x,k_y,k_z)=H^\ast(k_x,k_y,-k_z)$, where $H(\bk)$ is a 3D gapped Hamiltonian in the same basis. Here the sign before $k_z$ is inverted by $C_2*T$ because $C_2$ does not change $k_z$ and $T$ flips its sign. On $k_z=0$ and $k_z=\pi$ planes, the Hamiltonian is real. At each point on these planes, the real Hamiltonian along with the Fermi energy defines a projection from all $n+m$ bands (with real wavefunctions) to $n$ occupied bands, or equivalently, choosing $n$ independent vectors from an $n+m$-dimensional real vector space. Mathematically, the space of these projections is called real Grassmanian manifold, $G^R(m,m+n)$, the topology of which is well-known in algebraic topology\cite{Hatcher2002}. Specifically, its second homotopy group $\pi_2[G^R(m,m+n)]=Z_2$, which means the $k_z=0,\pi$ planes can be each associated with a $Z_2$ topological invariant. These two $Z_2$ indices give a $Z_2\times{Z}_2$ classification of the 3D bulk. How did we miss one $Z_2$ index in the analysis of the surface states? This is because on the surface $k_z$ is not a good quantum number, so the two indices cannot be probed distinctly on the surface. We claim that the strong $Z_2$ index, which indicates the presence/absence of the surface Dirac point is given by the sum of the two $Z_2$ indices from $k_z=0$ plane and $k_z=\pi$ plane $\delta_{3D}=\delta(k_z=0)+\delta(k_z=\pi)$, where $\delta(k_z=0/\pi)$ is the $Z_2$ index. The argument goes as follows: if the two planes have the same $Z_2$ indices, the 3D insulator is adiabatically equivalent to layers of decoupled 2D insulators, and cannot have gapless surface modes on the top layer which is insulating.

We can write down a simple spinless tight-binding model that shows a nontrivial $Z_2$ phase protected by $C_2*T$:
\bea\label{eq:TB2}\nonumber
&&H(\bk)=[m-\cos(k_x-a)-\cos(k_y-b)-\cos{k_z}]\Sigma_{03}\\
&&+\sin(k_x-a)\Sigma_{11}+\sin(k_y-b)\Sigma_{31}+\sin{k_z}\Sigma_{21},
\eea
where $2<m<3$ and $a,b$ are arbitrary numbers. It is straightforward to check that the Hamiltonian is invariant under $C_2*T=K$, i.e.,
\bea
H(k_x,k_y,k_z)=H^\ast(k_x,k_y,k_z).
\eea
The model of Eq.(\ref{eq:TB2}) shows a single Dirac cone at $(a,b)$ in the SBZ of the $(001)$-surface.


\end{document}